\documentclass[bibyear]{aa}

\usepackage{graphicx}

\usepackage{txfonts}

\usepackage{natbib}
\bibpunct{(}{)}{;}{a}{}{,}
\usepackage{booktabs}
\usepackage{multicol}
\usepackage{graphicx}
\usepackage{fancyhdr}

\usepackage{amsmath}
\usepackage{amssymb}
\usepackage[inter-unit-product=\cdot]{siunitx}
\usepackage{enumitem}
\usepackage{multirow}

\usepackage{mathtools}

\usepackage{longtable}
\usepackage{tabularx}
\usepackage{xcolor}
\usepackage{ulem}
\usepackage{comment}
\usepackage{cuted}

\usepackage{soul}

\usepackage{placeins}
\usepackage{stfloats}
\usepackage{float}

\usepackage{hyperref}
\usepackage{bm}

\hypersetup{colorlinks=true,urlcolor=black, citecolor=blue, linkcolor=black}

\makeatletter
\renewcommand*\aa@pageof{, page \thepage{} of \pageref*{LastPage}}
\makeatother

\newcommand{\orcidicon}[1]{\href{https://orcid.org/#1}{\includegraphics[width=11pt]{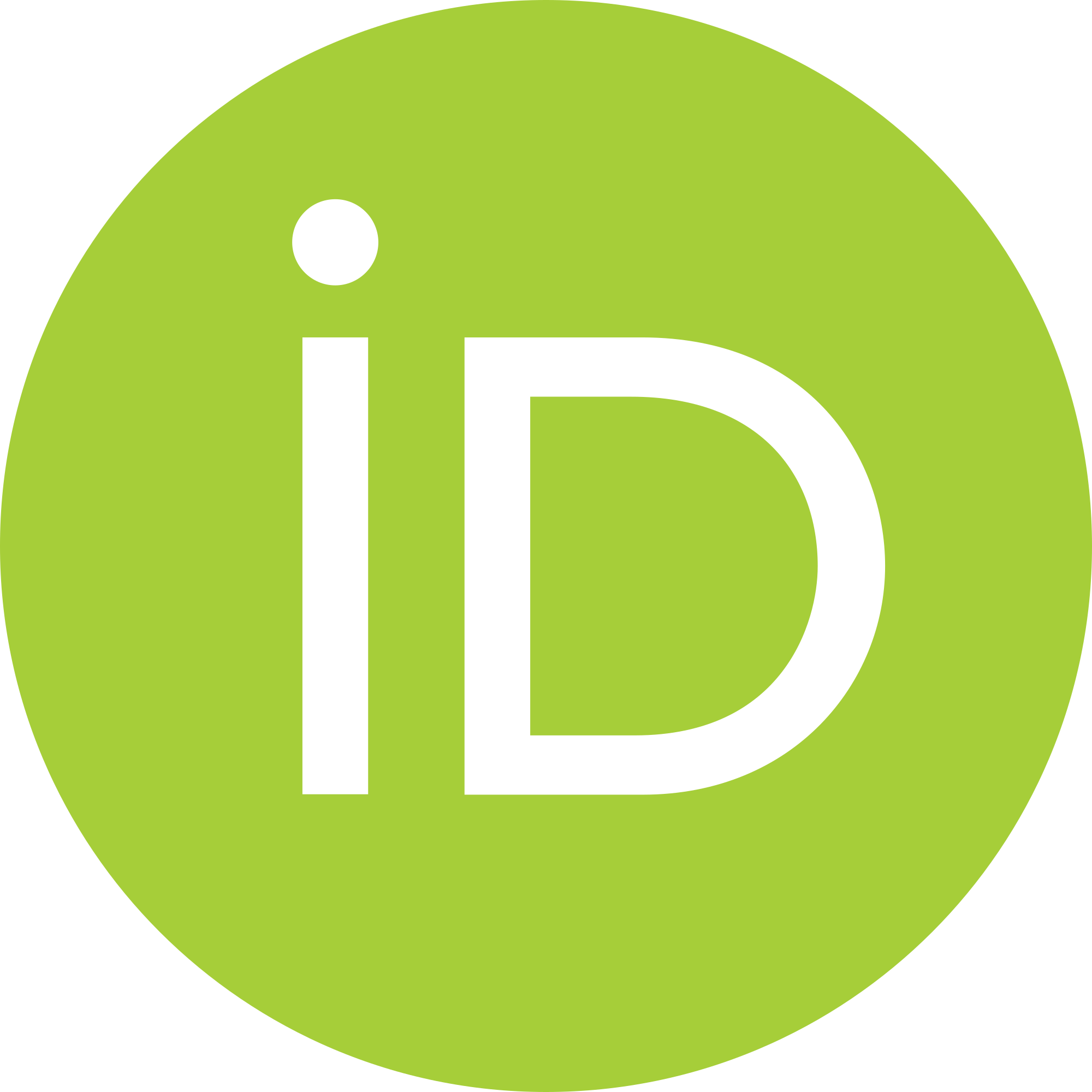}}}
\newcommand{\orcid}[1]{\href{https://orcid.org/#1}{\protect\orcidicon{#1}}}

\begin{document}

\title{Shaping binary black hole merger efficiency with gravitational~wave observations}
\titlerunning{}

\author{
Michele Bosi\thanks{\href{mailto:mbosi@sissa.it}{mbosi@sissa.it}}
\inst{1,2}
\orcid{0009-0000-8215-6698},
Lumen Boco\inst{3}
\orcid{0000-0003-3127-922X},
Stefano Torniamenti\inst{4}
\orcid{0000-0002-9499-1022},
Stefano Rinaldi\inst{3}
\orcid{0000-0001-5799-4155},
Cecilia Sgalletta\inst{3}
\orcid{0009-0003-7951-4820},
Michela~Mapelli\inst{3,5,6,7}
\orcid{0000-0001-8799-2548},
Carlo Baccigalupi\inst{1,8,9}
\orcid{0000-0002-8211-1630} ,
Andrea Lapi\inst{1,8,9,10}
\orcid{0000-0002-4882-1735}
}
\authorrunning{M. Bosi et al.}
\institute{
    $^{1}$Scuola Internazionale Superiore di Studi Avanzati, Via Bonomea 265, 34136 Trieste, Italy\\
    $^{2}$Department of Physics, University of Trento, Via Sommarive 14, 38123 Povo (TN), Italy\\
    $^{3}$Universit\"at Heidelberg, Zentrum f\"ur Astronomie (ZAH), Institut f\"ur Theoretische Astrophysik, Albert Ueberle Str. 2, 69120, Heidelberg, Germany\\
    $^{4}$Max-Planck-Institut für Astronomie, Königstuhl 17, 69117, Heidelberg, Germany\\
    $^{5}$Universit\"at Heidelberg, Interdiszipli\"ares Zentrum f\"ur Wissenschaftliches Rechnen, D-69120 Heidelberg, Germany\\
    $^{6}$Dipartimento di Fisica e Astronomia Galileo Galilei, Università di Padova, Vicolo dell’Osservatorio 3, I–35122 Padova, Italy\\
    $^{7}$INFN, Sezione di Padova, Via Marzolo 8, I--35131 Padova, Italy\\
    $^{8}$Institute for Fundamental Physics of the Universe (IFPU), Via Beirut 2, 34014 Trieste, Italy\\
    $^{9}$Istituto Nazionale Fisica Nucleare (INFN), Sezione di Trieste, Via Valerio 2, 34127 Trieste, Italy\\
    $^{10}$INAF - Istituto di Radioastronomia, Via Gobetti 101, 40129 Bologna, Italy
    }

\date{Received XXXX; accepted YYYY}

\abstract{Gravitational wave (GW) astronomy offers unprecedented insights into binary black hole (BBH) coalescence.
However, many of the key quantities involved remain inaccessible to direct observations.
The BBH merger rate density is deeply linked to both the merger efficiency and the distribution of delay time between binary formation and merger, neither of which is directly constrained by observations.
Disentangling their impact on the merger rate represents a highly non-trivial endeavour.
Here we present a semi-parametric BBH population model, based on population synthesis simulations of both isolated and dynamically-formed BBHs, anchored on an observation-driven, metallicity-dependent star formation history.
We parametrise the merger efficiency of these two formation channels, fitting the model to GW events from the Gravitational-Wave Transient Catalog 5.0 within a hierarchical Bayesian framework.
Our analysis suggests that the isolated BBH merger efficiency should be lowered by a factor~$\mathcal{O}(10)$ relative to standard population synthesis results.
The dynamical channel requires an efficiency more than an order of magnitude larger than the isolated one to reproduce current GW observations.
Nevertheless, we find that the two channels provide comparable contributions to the observed number of events.
Finally, we introduce a parametrisation for the delay time distribution of the isolated BBHs.
We derive a distribution consistent with the observed local merger rate and show its degeneracy with the merger efficiency.
}

\keywords{Gravitational Waves – Black hole physics – Stars: black holes - globular clusters: general - Galaxies: star formation}

   \maketitle

\section{Introduction}\label{sec:intro}
In the past decade binary black hole (BBH) astrophysics has been thoroughly investigated by gravitational wave astronomy.
The growing number of GW events detected by the LIGO~\citep{Aasi2015:ligo}, Virgo~\citep{Acernese2015:virgo} and KAGRA~\citep{Akutsu2021:kagra} observatories (LVK), across the O1~\citep{Abbott2016:o1}, O2~\citep{Abbott2019:o2}, O3~\citep{Abbott2021b:o3_1, Abbott2023b:o3_3, Abbott2024:o3_2} and O4~\citep{Abac2025:o4_1,Abac2026:o4_2} observing runs,   enables statistical inference on the properties of coalescing compact object populations.
Features in the BBH mass spectrum are assessed with both phenomenological~\citep{Talbot2018:mass,Wysocki2019:phen} or more agnostic~\citep{Rinaldi2021:mass,Callister2023:non_par,Heinzel2024:non_par,Heinzel2025:non_par,Tenorio2025:non_par,Afroz2026,Gennari2026:mass_spectrum} methodologies.
They carry crucial implications for processes driving binary evolution~\citep[e.g.][]{Golomb2024:astro,Mould2026:mass} and may reveal whether BBH masses evolve with redshift;
though, at the present epoch, the mass-redshift correlation is non-trivial to establish given the limited redshift range spanned by current observations~\citep{Fishbach2021:red_evol,Gennari2025:spsirens,Lalleman2025:red_evol, Rinaldi2025:red_evol,Cheng2026:red_evol}.
Furthermore, the BBH mass spectrum provides clues on contributions from different formation channels~\citep{Zevin2021, Gerosa2021:hm, Cheng2023, Colloms2025, Galaudage2026:fc}, a picture that can be further probed by inspecting the mass-spin correlation~\citep[e.g.][]{Sadiq2025:spin,Berti2025:spin,Biscoveanu2026:spin} and the GW anisotropic distribution~\citep[e.g.][]{Chakravarti2026,Bellomo2026}.

Although GW population inference has significantly narrowed the parameter space of BBH coalescence, many underlying astrophysical processes cannot be directly assessed from detected events and require accurate modeling to provide robust predictions.
The comparison of the theoretical BBH merger rate density with that inferred by the LVK collaboration is a typical and fruitful testbed for the uncertainties affecting processes of this sort.
The BBH merger rate density is usually obtained by populating cosmic epochs with outputs of binary evolution simulations throughout a metallicity-dependent star formation history model.
The evolution of isolated binaries is studied through binary population synthesis codes such as~\textsc{seba}~\citep{Zwart1996:seba},~\textsc{bse}~\citep{Hurley2002:bse},~\textsc{sevn}~\citep{Spera2017, Spera2019, Mapelli2020, Iorio2023},~\textsc{mobse}~\citep{Mapelli2017,Giacobbo2017},~\textsc{compas}~\citep{Stevenson2017:compas,Barrett2018:compas,VignaGomez2018:compas,Broekgaarden2019:compas,Neijssel2019},~\textsc{bpass}~\citep{Eldridge2017:bpass,Stanway2018:bpass,Byrne2022:bpass},~\textsc{combine}~\citep{Kruckow2018:combine},~\textsc{startrack}~\citep{Chruslinska2019a, Belczynski2020},~\textsc{cosmic}~\citep{Breivik2019:cosmic},~\textsc{metisse}~\citep{Agrawal2020:metisse},~\textsc{posydon}~\citep{Fragos2022:pos,Andrews2024:pos} and~\textsc{binary\_c}~\citep{Hendriks2023:binaryc}.
For dynamical formation channels, the formation and evolution of BBHs in massive star clusters is being explored by a new class of semi-analytic codes like~\textsc{cbhbd}~\citep{Antonini2019:cbhbd,Antonini2023,Pouliasis2026:cbhbd},~\textsc{b-pop}~\citep{Sedda2020:bpop,Sedda2021:bpop,Sedda2026:bpop},~\textsc{fastcluster}~\citep{Mapelli2021,Mapelli2022,Vaccaro2023,Torniamenti2024} and~\textsc{rapster}~\citep{Kritos2023:rap,Kritos2024:rap}.

The cosmic star formation rate and metallicity distribution are modeled either through cosmological simulations~\citep{Mapelli2017,OShaughnessy2017,Schneider2017,Lamberts2018,Mapelli2018,Artale2019,Briel2023,Levina2026} or via observation-driven relations among the host galaxy properties, as star formation rate, stellar mass and metallicity~\citep{Dominik2013,Belczynski2016, Lamberts2016, Cao2018,Elbert2018,Li2018,Boco2019,Chruslinska2019,Neijssel2019,Santoliquido2020,Tang2020,Olejak2021,Boco2021,Broekgaarden2021,Santoliquido2022,Broekgaarden2022,Romagnolo2023:mr,Iorio2023,Romagnolo2025:mr,Sgalletta2025,Boco2026a,Boco2026b}.

Direct comparisons between the observationally inferred merger rate density and theoretical models combining binary evolution simulations with star formation history can reveal a known tension, as predicted merger rates might mismatch observations depending on the underlying assumptions~\cite[e.g.][]{Broekgaarden2022,Santoliquido2022, Srinivasan2023:merger_rate, Boesky2024:merger_rate,Sgalletta2025,Boco2026a,Broekgaarden2026:mr,Boco2026b}.
To investigate this discrepancy, we adopt an inverse approach (see e.g.~\cite{Rauf2024:ba} for a similar procedure).
Instead of constructing a theoretical merger rate density totally anchored on the results of binary population synthesis, we parametrise key outputs of the simulations and couple them with an observation-based star formation history.
We then fit this BBH population model to GW data in order to infer the parameter values required to reconcile theoretical predictions with current observations.
Among the outputs of population synthesis codes, the quantities that most strongly determine the predicted merger rate density are the BBH merger efficiency (i.e. the number of merging BBHs within a Hubble time per unit of simulated stellar mass, at a given metallicity) and the delay time distribution (i.e. the time between the binary formation and merger).
Given that both these quantities are subject to substantial theoretical uncertainties, our goal is to constrain them directly from observed GW events.

Following this line, we present a semi-parametric BBH population model based on an observation-driven, metallicity-dependent star formation history.
We implement a multi-formation channel scenario, simulating both isolated BBHs and dynamically-formed BBHs in globular clusters (GCs) to cover the redshift and mass ranges probed by current GW observations~\citep{lvk2026:pop}.
By parametrising the merger efficiency of these two distinct formation channels, we infer their values fitting the model to GW events from the Gravitational-Wave Transient Catalog 5.0 (GWTC-5.0)~\citep{Abac2026:o4_2}, within a full hierarchical Bayesian framework~\citep{Mandel2019,Vitale2020}.
This framework enables us to account properly for the observed number of events alongside their redshift and mass dependence.
We estimate then the contribution of the two channels to the total BBH merger rate density and mass distribution, as well as their relative abundance.
Furthermore, we introduce a parametrisation for the delay time distribution of isolated BBHs and, employing the same inference framework, we study its effect on the redshift evolution of the merger rate density.
Lastly, since changes in both delay times and merger efficiencies might alter the number of BBHs in the local Universe, we investigate how they correlate in shaping the overall cosmic merger rate density.

The paper is organised as follows: in Section~\ref{sec:stat_frame} we set the statistical framework of the analysis.
In Section~\ref{sec:pop_model} we describe the BBH population model, starting from the binary simulations, in Section~\ref{sec:pop_synt}, moving to the star formation history model, in Section~\ref{sec:sfr_hist}, and to the parametric approach in Section~\ref{sec:merger_rate model}.
In Section~\ref{sec:results} we present our results, in Section~\ref{sec:discussion} we discuss the main caveats of our approach and, finally, we conclude in Section~\ref{sec:conclusions}.
The Appendices provide additional information about the GW dataset (\ref{sec:appendix_gw_data}), the inference procedure (\ref{sec:appendix_likelihood}), modeling (\ref{sec:appendix_bbh_cat} and \ref{sec:appendix_csfrd}) and final results (\ref{sec:appendix_model_td}).

Throughout the paper we assume Planck 2015~\citep{Planck2015:cosmo} flat~$\Lambda$ cold dark matter (CDM) cosmolgy with~$H_0\simeq67.90$,~$\Omega_m\simeq0.31$ and~$\Omega_\Lambda\simeq0.69$.
Concerning metallicity, the reference solar abundances are~$12+\log(\textrm{O}/\textrm{H})_\odot=8.83$,~$12+\log(\textrm{Fe}/\textrm{H})_\odot=7.5$,~$\log(\textrm{O}/\textrm{Fe})_\odot=1.33$, and~$Z_\odot=0.017$, for consistency with~\citep{Grevesse1998,Chruslinska2025, Boco2026a}.

\section{Statistical framework}\label{sec:stat_frame}
We adopt a hierarchical Bayesian framework to compare our BBH merger rate density model with GW data.
Such statistical framework allows to infer population parameters from the posterior samples of collected GW observations~\citep{Loredo1995,Mandel2019,Vitale2020}.
Specifically, the hierarchical likelihood of observing the strain data associated to~$N_\mathrm{obs}$ GW events,~$\bm{x}=\{x_1,...,x_{N_\mathrm{obs}}\}$, within an observation time~$T_\mathrm{obs}$, and for a given set of hyperparameters~$\bm{\Lambda}$, is~\citep[e.g.][]{Loredo2002,Farr2015,Thrane2019}:
\begin{equation}
    \mathcal{L}(\bm{x}|\bm{\Lambda})\propto\ e^{-N_\mathrm{exp}(\bm{\Lambda})}\prod^{N_\mathrm{obs}}_iT_\mathrm{obs}\int \mathrm{d}\bm{\theta}\mathrm{d}z\ \mathcal{L}_\mathrm{ev}(x_i|\bm{\theta},z,\bm{\Lambda}_c)\frac{\mathrm{d}N_\mathrm{BBH}}{\mathrm{d}\bm{\theta}\mathrm{d}z\mathrm{d}t_d}(\bm{\Lambda}),
\label{eq:likelihood}
\end{equation}
where~$z$ is the redshift,~$\bm{\theta}$ are the single-event parameters (e.g. masses and spins of the binary) and ~$\bm{\Lambda}=\{\bm{\Lambda}_c,\bm{\Lambda}_p\}$ includes both the cosmological,~$\bm{\Lambda}_c$, and population,~$\bm{\Lambda}_p$, parameters.
~$\mathcal{L}_\mathrm{ev}(x_i|\bm{\theta},z,\bm{\Lambda}_c)$ is the single-event likelihood and~$\mathrm{d}N_\mathrm{BBH}/\mathrm{d}\bm{\theta}/\mathrm{d}z\mathrm/{d}t_d$ is the differential BBH merger rate in the detector-frame time~$t_d$, i.e. the population model.
~$N_\mathrm{exp}(\bm{\Lambda})$ is the expected number of GW events by the population model itself, it takes into account selection effects due to the detector~\citep{Loredo2002,Gaebel2019,Vitale2020} and can be computed as:
\begin{equation}
    N_\mathrm{exp}(\bm{\Lambda}) = T_\mathrm{obs}\int \mathrm{d}\bm{\theta}\mathrm{d}z\ P_\mathrm{det}(\bm{\theta},z,\bm{\Lambda}_c)\frac{\mathrm{d}N_\mathrm{BBH}}{\mathrm{d}\bm{\theta}\mathrm{d}z\mathrm{d}t_d}(\bm{\Lambda}),
    \label{eq:Nexp}
\end{equation}
with~$P_\mathrm{det}(\bm{\theta},z,\bm{\Lambda}_c)$ the detection probability that we compute as in equation (6) of~\cite{Mastrogiovanni2023:icarogw}.

The differential BBH merger rate can be expressed as:
\begin{equation}
    \frac{\mathrm{d}N_\mathrm{BBH}}{\mathrm{d}\bm{\theta}\mathrm{d}z\mathrm{d}t_d} = \frac{1}{1+z}\frac{\mathrm{d}N_\mathrm{BBH}}{\mathrm{d}\bm{\theta}\mathrm{d}t_s\mathrm{d}V_c}\frac{\mathrm{d}V_c}{\mathrm{d}z}=\frac{1}{1+z}\frac{\mathrm{d}\mathcal{R}_\mathrm{BBH}}{\mathrm{d}\bm{\theta}}\frac{\mathrm{d}V_c}{\mathrm{d}z},
\end{equation}
with~$t_s$ the source-frame time,~$\mathrm{d}V_c/\mathrm{d}z$ the comoving volume element and~$\mathrm{d}\mathcal{R}_\mathrm{BBH}/\mathrm{d}\bm{\theta}$ the differential BBH merger rate density in source-frame time.
This is the key quantity where we embed our semi-parametric model based on BBH population synthesis simulations, introduced in Section~\ref{sec:pop_model}.
In this analysis, we consider only the source-frame BBH primary mass,~$m_1$, the mass ratio,~$q=m_2/m_1$, and the redshift among the single-event parameters, leaving the inclusion of spin information for future development.

We analyse BBH events from the GWTC-5.0~\citep{Abac2026:o4_2} with the~\textsc{Python} packages~\textsc{figaro}\footnote{\textsc{figaro} is publicly available at \url{https://github.com/sterinaldi/FIGARO}.}~\citep{Rinaldi2024:figaro} for data processing, and~\textsc{icarogw}\footnote{\textsc{icarogw} is publicly available at \url{https://github.com/simone-mastrogiovanni/icarogw}.}~\citep{Mastrogiovanni2023:icarogw} for likelihood evaluation.
More details on the analysed GW events and likelihood computation can be found in Appendix~\ref{sec:appendix_gw_data} and~\ref{sec:appendix_likelihood}, respectively.

\section{BBH population model}\label{sec:pop_model}
As previously mentioned, the BBH merger rate density can be obtained as the product of the interplay between a metallicity-dependent cosmic star formation history and binary evolution.
Hence, in the first place, we simulate populations of BBHs originating in different environments.
Then we derive a metallicity-dependent cosmic star formation rate density (SFRD) from recent observations.
Finally, we assemble these ingredients in a semi-parametric BBH merger rate density model that we exploit in the inference procedure.
The modeling is described below.

\subsection{Population synthesis}\label{sec:pop_synt}
Current GW data suggest that BBH mergers could have multiple formation channels relative to specific astrophysical environments~\citep[e.g.][]{Zevin2017,Bouffanais2019,Bouffanais2021,Zevin2021, Cheng2023, Colloms2025, Galaudage2026:fc}.
Indeed, features in both the mass and spin distributions suggest the presence of BBH subpopulations~\citep[e.g.][]{Abbott2025b, lvk2026:pop}.
Thus, we simulate two populations of BBHs, corresponding to two formation channels: isolated BBHs and dynamically-formed BBHs in globular clusters.
We use~\textsc{sevn}\footnote{We use the~\textsc{sevn} version V 2.16.0.
\textsc{sevn} is publicly available at \url{https://gitlab.com/sevncodes/sevn}.}~\citep{Spera2017, Spera2019, Mapelli2020, Iorio2023} to generate catalogs of isolated BBHs resulting from the evolution of stellar binaries in the field.
We will refer to this formation scenario as the~``isolated formation channel'' (ISO).
In particular, we evolve a population of~$10^7$ binaries for~$15$ metallicity values ranging from~$Z_\mathrm{sim}=0.0002$ to ~$Z_\mathrm{sim}=0.02$.
Then, we use the code \textsc{fastcluster}\footnote{We adopt the \textsc{fastcluster} version developed in~\cite{Torniamenti2024}. The open-source version is publicly available at \url{https://gitlab.com/micmap/fastcluster_open}.}~\citep{Mapelli2021, Mapelli2022, Vaccaro2023, Torniamenti2024} to simulate the dynamical formation and evolution of~$10^6$ BBHs in GCs for the same metallicity values.
We will refer to this formation scenario as the~``globular cluster formation channel'' (GC).
Further details on the simulated BBH catalogs are reported in Appendix~\ref{sec:appendix_bbh_cat}.

For the purposes of this work, as a first approximation, we consider the GC formation channel as representative of the whole dynamical channel.
This is motivated by the fact that, as shown in previous studies~\citep{Antonini2023,Torniamenti2024}, the GC formation channel is capable of reproducing features of the high-mass tail of the BBH mass spectrum.
The inclusion of other formation scenarios will be part of future extensions of this analysis.

\begin{figure}
    \centering
    \includegraphics[width=0.9\linewidth]{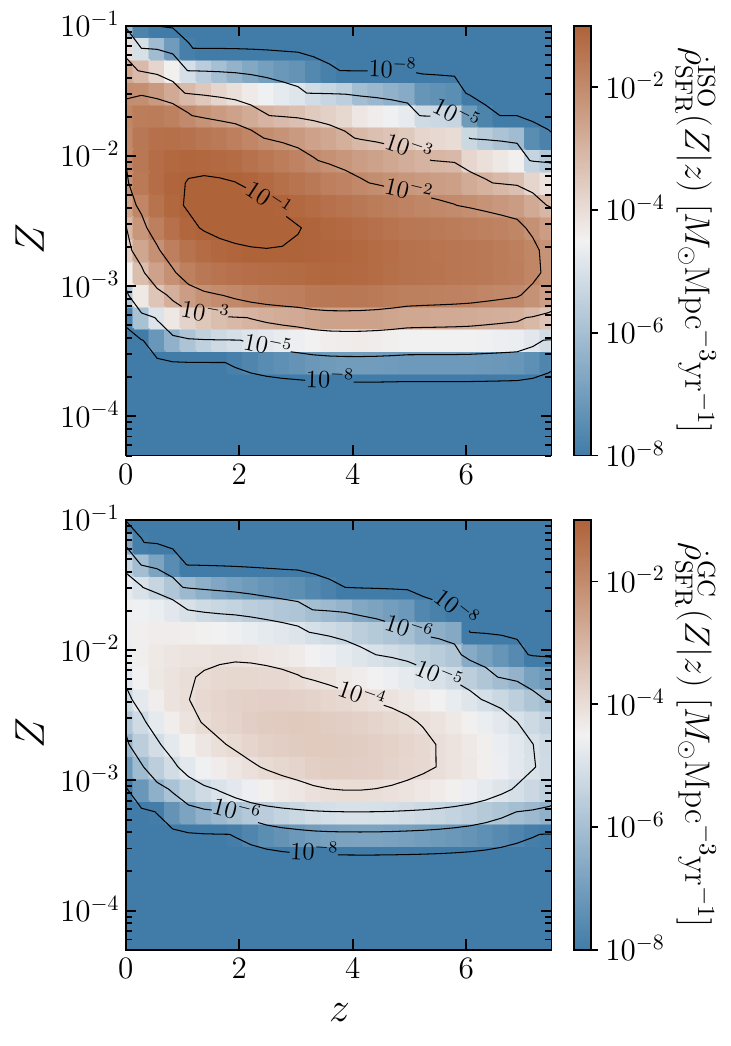}
    \caption{Cosmic star formation rate density feeding the isolated (\textit{top panel}) and globular cluster (\textit{bottom panel}) BBH formation channel, as function of both redshift and metallicity.
    The metallicity dependence is guided by the iron-group elements' abundance, as described in Appendix~\ref{sec:appendix_csfrd}.
    The black contours show levels of constant cosmic SFRD ranging from~$10^{-8}$ to~$10^{-1}$ M$_\odot$Mpc$^{-3}$yr$^{-1}$(\textit{top panel}) and~$10^{-4}$  M$_\odot$Mpc$^{-3}$yr$^{-1}$ (\textit{bottom panel}).}
    \label{fig:csfrd_Z}
\end{figure}

\subsection{Star formation history}\label{sec:sfr_hist}

To compute the metallicity-dependent total cosmic SFRD, we adopt the observation-based approach of~\cite{Boco2021}.
We calculate it as:
\begin{equation}
    \begin{aligned}
        \dot{\rho}^\mathrm{tot}_\mathrm{SFR}(Z|z) &= \int \mathrm{d}M_\star \frac{\mathrm{d}N_\mathrm{gal}}{\mathrm{d}M_\star\mathrm{d}V_c}(M_\star|z)\ \times\\
        &\quad \times \int \mathrm{d}\mathrm{SFR} p(\mathrm{SFR}|M_\star,z)p(Z|M_\star,\mathrm{SFR})\mathrm{SFR},
    \end{aligned}
    \label{eq:csfrd}
\end{equation}
where~$\mathrm{d}N_\mathrm{gal}/\mathrm{d}M_\star/\mathrm{d}V_c$ is the star-forming galaxy stellar mass function,~$p(\mathrm{SFR}|M_\star,z)$ is the conditional probability density function (pdf) of the star formation rate (SFR) at given stellar mass and redshift, and~$p(Z|M_\star,\mathrm{SFR})$ is the metallicity conditional pdf, depending on both stellar mass and SFR.
The adopted distributions are described in Appendix~\ref{sec:appendix_csfrd}.

\subsubsection{Multi-channel approach}\label{sec:sfr_hist_mc}
Next, we calculate the amount of star formation occurring in GCs and in the field, to estimate the contribution of the two channels to the total merger rate density.
Assuming that the same metallicity dependence applies for both the channels, the cosmic SFRD at a given metallicity value in GCs can be expressed as
\begin{equation}
    \dot{\rho}^\mathrm{GC}_\mathrm{SFR}(Z|z) = \mathcal{F}_\mathrm{SFR}(z)\ \dot{\rho}^\mathrm{tot}_\mathrm{SFR}(Z|z),
\end{equation}
where~$\mathcal{F}_\mathrm{SFR}(z) \equiv \dot{\rho}^\mathrm{GC}_\mathrm{SFR}(z)/ \dot{\rho}^\mathrm{tot}_\mathrm{SFR}(z)$, with~$\dot{\rho}^\mathrm{tot}_\mathrm{SFR}$ given by equation~\eqref{eq:csfrd} marginalised over~$Z$.
$\dot{\rho}^\mathrm{GC}_\mathrm{SFR}(z)$ is the GC star formation rate density as a function of redshift only, assumed to follow a Gaussian distribution as:
\begin{equation}
    \dot{\rho}^\mathrm{GC}_\mathrm{SFR}(z) = \mathcal{B}_\mathrm{GC}\ \mathrm{exp}\left[{-\frac{(z-z_\mathrm{GC})^2}{2\sigma_\mathrm{GC}^2}}\right],
    \label{eq:csfrd_gc}
\end{equation}
with~$\sigma_\mathrm{GC}=1.5$ and~$z_\mathrm{GC}=3.2$~\citep{Mapelli2022} and~$\mathcal{B}_\mathrm{GC}=2\times10^{-4}\ \mathrm{M}_\odot\mathrm{Mpc}^{-3}\mathrm{yr}^{-1}$ in agreement with~\cite{rodriguez2018, elbadry2019,ReinaCampos2019:gc}.
Thereby, the fraction of cosmic SFRD relative to the isolated channel is given by
\begin{equation}
    \dot{\rho}^\mathrm{ISO}_\mathrm{SFR} = \dot{\rho}^\mathrm{tot}_\mathrm{SFR} - \dot{\rho}^\mathrm{GC}_\mathrm{SFR}.
\end{equation}
Figure~\ref{fig:csfrd_Z} shows the two contributions varying with both redshift and metallicity. It can be noticed how the star formation happening in the field is dominant by orders of magnitude across all the redshift range.

\subsection{Semi-parametric merger rate density}\label{sec:merger_rate model}
Within our multi-channel framework, the total differential BBH merger rate density per unit of the BBH primary mass in the source-frame and mass ratio can be written as:
\begin{equation}
    \frac{\mathrm{d}\mathcal{R}^\mathrm{tot}_\mathrm{BBH}}{\mathrm{d}m_1\mathrm{d}q}= \frac{\mathrm{d}\mathcal{R}^\mathrm{ISO}_\mathrm{BBH}}{\mathrm{d}m_1\mathrm{d}q}+\frac{\mathrm{d}\mathcal{R}^\mathrm{GC}_\mathrm{BBH}}{\mathrm{d}m_1\mathrm{d}q},
    \label{eq:mergerRateDensity}
\end{equation}
where~$\mathrm{d}\mathcal{R}^i_\mathrm{BBH}/\mathrm{d}m_1/\mathrm{d}q$ with~$i=\mathrm{ISO}\ (\mathrm{GC})$ stands for the contribution of the ISO (GC) formation channel.
We derive the differential merger rate density of each single channel as~\citep[e.g.][]{Boco2021, Broekgaarden2022}:
\begin{equation}
    \begin{aligned}
        \frac{\mathrm{d}\mathcal{R}^{i}_\mathrm{BBH}}{\mathrm{d}m_1\mathrm{d}q} &=\int\mathrm{d}Z\ \eta_i(Z)p_i(m_1,q|Z)\ \times\\
        &\quad\times\int^{t(z)}_{t^\mathrm{min}_d}
        \mathrm{d}t_d\ p_i(t_d|Z)\dot{\rho}^i_\mathrm{SFR}(Z|t(z)-t_d),
    \end{aligned}
    \label{eq:mergerRateDensity_i}
\end{equation}
where~$t_d$ is the delay time between binary star formation and BBH merger,~$t(z)$ is the cosmic time at the redshift of the binary merger,~$z$, and~$\dot{\rho}^i_\mathrm{SFR}(Z|t(z)-t_d)$ is the cosmic SFRD of the~$i$-th channel, calculated at the cosmic time of binary formation,~$t(z_f)=t(z)-t_d$.
Next, the cosmic SFRD is weighted with a metallicity-dependent pdf of the delay time,~$p_i(t_d|Z)$, and a joint pdf, still conditional on metallicity, of both the primary mass and mass ratio,~$p_i(m_1,q|Z)$.
While the metallicity-dependent cosmic SFRD is built as described in Section~\ref{sec:sfr_hist_mc}, the two distributions~$p_i(t_d|Z)$ and~$p_i(m_1,q|Z)$ are directly extracted from the output of the population synthesis codes (at least in the first part of this work)~\footnote{As a first approximation, we consider~$p_i(m_1,q|Z)$ and~$p_i(t_d|Z)$ to be independent.
However,~$\{m_1,q,t_d\}$ are inherently correlated through binary interaction processes.
While adopting a joint pdf is computationally more expensive, it would prevent biases in the redshift evolution of the merger rate density, especially at high redshifts, as shown in~\cite{Antinozzi2026}.}.
Finally,~$\eta_i(Z)$ is the merger efficiency of the~$i$-th formation channel, which provides the number of BBH mergers per unit of the total simulated initial stellar mass, at a given metallicity.

In the case of isolated binaries, many studies~\citep[e.g.][]{Giacobbo2018, Klencki2018, Neijssel2019, Iorio2023} have found that the merger efficiency has a strong metallicity dependence, taking values~$\sim 10^{-4.5}$ at~$Z\lesssim Z_\odot/3$ and featuring a huge drop (3-4 orders of magnitudes) at higher~$Z$.
This behavior is mostly determined by the strength of stellar winds.
However, both the value of the efficiency at low-$Z$ and the exact position of the drop are rather uncertain and depend on numerous aspects~\citep{Vanson2025}, ranging from initial conditions, single stellar evolution (stellar winds, supernova explosion mechanism, natal kicks) and binary evolution (mass transfer and common envelope formalism).
As for GCs, simulations do not predict a clear metallicity dependence of the BBH merger efficiency, as shown in Figure 5 of~\cite{Rodriguez2016} for GCs, but also in Figure 9 of~\cite{DiCarlo2019} for young star clusters.
The value of the efficiency depends on several factors, such as the initial binary fraction in the clusters, their initial mass and density profile and the natal kick distribution~\citep[e.g.][]{Rodriguez2016,Chatterjee2017,Hong2018}.

The merger efficiency is one of the primary elements that regulate the overall normalisation and redshift evolution of the BBH merger rate density. Recently, \cite{Sgalletta2025, Boco2026a} showed that the BBH merger rate density predicted by theoretical models can be overestimated with respect to observations, depending on the underlying assumptions.
They also show that this overestimation cannot be reconciled by changing the galaxy main sequence~\citep{Sgalletta2025} or metallicity relations~\citep{Boco2026a}.
Therefore, they suggest that this discrepancy stems mainly from outcomes of population synthesis models rather than from the treatment of the cosmic SFRD.

The goal of this work is to estimate the value and metallicity dependence that the BBH merger efficiency should have to reconcile with GW data.
Given the high dimensionality of the problem, with the merger efficiency depending on a plethora of astrophysical processes, we choose to follow a parametric approach.
Namely, we parametrise the merger efficiency of the two formation channels as:
\begin{equation}
    \begin{aligned}
        &\eta_\mathrm{ISO}(Z)=A_\mathrm{ISO}\Theta_H(\mathrm{log}_{10}Z_\mathrm{max}-\mathrm{log}_{10}Z)\\
        &\eta_\mathrm{GC}(Z)=A_\mathrm{GC},
    \end{aligned}
    \label{eq:efficiency}
\end{equation}
where~$A_\mathrm{ISO}$ and~$A_\mathrm{GC}$ are constant amplitudes in metallicity and~$Z_\mathrm{max}$ is the cut-off metallicity.
In addition, we define the ratio of the two efficiency amplitudes
\begin{equation}
    f_\mathrm{GC} = A_\mathrm{GC} / A_\mathrm{ISO}.
    \label{eq:eff_amp_ratio}
\end{equation}
We estimate the merger efficiency parameters by fitting GW events from the GWTC-5.0 via a hierarchical Bayesian framework. In doing so, we reconstruct the values that these parameters should take to match the predicted and observed BBH merger rate density and mass function.
This approach is agnostic about which of the aforementioned astrophysical processes would lead to an adjustment of the merger efficiency, but it provides data-calibrated constraints on the values of the efficiency parameters.

\section{Results}\label{sec:results}
Here, we present the results of the hierarchical Bayesian inference on BBH events from the GWTC-5.0 catalog with our semi-parametric approach.

First, we present constraints on the merger efficiency from the BBH population model with the efficiencies parametrised as in equation~\eqref{eq:efficiency}, while delay time and mass pdfs are derived from simulations.
In the following, we refer to this model as the~$\{Z_\mathrm{max},A_\mathrm{ISO},f_\mathrm{GC}\}$ model.
Second, we discuss how the results may be affected by including a power-law parametrisation for the delay time pdf in the isolated formation scenario.
Such model is labeled as the~$\{Z_\mathrm{max},A_\mathrm{ISO},f_\mathrm{GC},\alpha_{d}\}$ model, with~$\alpha_d$ the exponent of the new delay time pdf.

\subsection{BBH merger efficiency}
\label{sec:mrd_model1}

\begin{figure}
    \centering
    \includegraphics[width=1.0\linewidth]{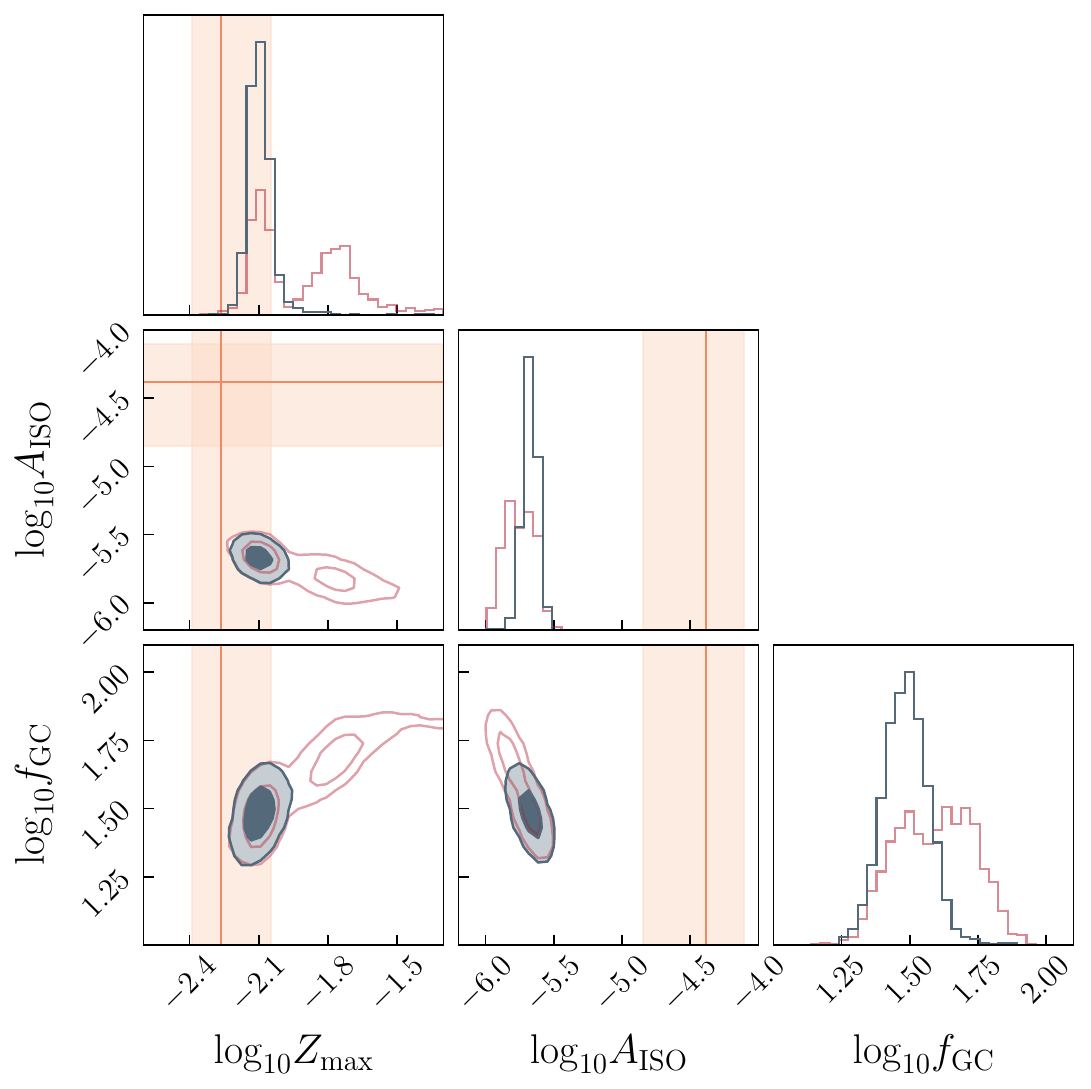}
    \caption{Corner plot with the marginalised posterior distributions of the~$\{Z_\mathrm{max},A_\mathrm{ISO},f_\mathrm{GC}\}$ model.
    The unfilled dark red inner (outer) contour encloses the~$50\%$ ($90\%$) credible region obtained with the complete GW dataset.
    The filled dark (light) blue contour corresponds to the~$50\%$ ($90\%$) credible region derived with the reduced GW dataset.
    The orange solid lines (shaded bands) denote median values ($80\%$ credible intervals) derived from Figure 1 of~\cite{Vanson2025} for qualitative comparison.}
    \label{fig:merger_eff_cp}
\end{figure}

To find the merger efficiencies that yield a merger rate density and mass distribution consistent with LVK estimates, we treat~$\mathrm{log}_{10}Z_\mathrm{max}$,~$\mathrm{log}_{10}A_\mathrm{ISO}$, and ~$\mathrm{log}_{10}f_\mathrm{GC}$, as free parameters of our inference procedure.
We show the relative posteriors in Figure~\ref{fig:merger_eff_cp}.

The dark red contours represent the marginalised distributions obtained considering all the observed 259 BBH mergers with false alarm rate~$<1\ \mathrm{yr}^{-1}$ (hereafter the ``complete GW dataset''), i.e.\ the same selection adopted in~\cite{lvk2026:pop}.
There is a clear bimodality in the posterior distributions.
In particular, there is one solution that peaks at~$\mathrm{log}_{10}Z_\mathrm{max}\simeq -2.1$, in agreement with what most population synthesis models predict, and a secondary solution that peaks at~$\mathrm{log}_{10}Z_\mathrm{max}\simeq-1.8$.
However, we found that the solution at high~$Z_\mathrm{max}$ is mainly due to three highly asymmetric binaries with~$q<0.5$, of which GW241011\_233834~\citep{lvk2025:GW241011} is the most significant.
Within our model, such low mass ratios could be reproduced by isolated binaries at high metallicity (Figure~\ref{fig:NBBH_sevn}). Thus, the fit favours a higher maximum metallicity~$Z_\mathrm{max}$ to accomodate these few highly asymmetric events.
However, such systems can be naturally interpreted as the result of hierarchical mergers or dynamical processes in AGN disks~\citep[e.g.][]{Vaccaro2023,Vaccaro2026:agn,Torniamenti2026b}, but their origin is still debated~\citep[e.g.][]{lvk2025:GW241011,Garcia2026:q}.
Since our aim is to characterise the bulk of the BBH population, in the rest of the analysis we exclude the events whose~$q$ lies below~$0.5$ with~$90\%$ probability, namely GW190412\_053044~\citep{lvk2020:GW190412}, GW240921\_201835~\citep{Abac2026:o4_2} and GW241011\_233834.
We verified that this exclusion does not affect appreciably the inferred merger rate density.
Hereafter, we refer to this 256-event sample as ``reduced GW dataset''.
\begin{table}[t]
   \centering
   \small
   \setlength{\tabcolsep}{3.5pt}
   \renewcommand{\arraystretch}{1.8}
   \setlength{\arrayrulewidth}{0.7pt}
   \begin{tabular}{lccc}
       \hline
       Parameter & Prior
       & $\{Z_\mathrm{max},A_\mathrm{ISO},f_\mathrm{GC}\}$
       & $\{Z_\mathrm{max},A_\mathrm{ISO},f_\mathrm{GC},\alpha_{d}\}$ \\
       \hline
       $\log_{10} Z_\mathrm{max}$ & $\mathcal{U}(-3.0,-1.3)$ & $-2.10^{+0.10}_{-0.07}$ & $-2.08^{+0.23}_{-0.07}$ \\

       $\log_{10} A_\mathrm{ISO}$ & $\mathcal{U}(-7.0,-3.0)$ & $-5.68^{+0.09}_{-0.09}$ & $-5.40^{+0.17}_{-0.34}$  \\

       $\log_{10} f_\mathrm{GC}$ & $\mathcal{U}(-1.0,3.0)$ & $1.49^{+0.13}_{-0.13}$ &~$1.21^{+0.37}_{-0.18}$  \\

       ~$\alpha_{d}$ & $\mathcal{U}(-2.0,0.0)$ & $-$ &~$-1.33^{+0.37}_{-0.51}$ \\

       \hline
   \end{tabular}
   \caption{Parameters (\textit{first column}), priors (\textit{second column}) and inferred constraints for the $\{Z_\mathrm{max}, A_\mathrm{ISO}, f_\mathrm{GC}\}$ (\textit{third column}) and $\{Z_\mathrm{max},A_\mathrm{ISO},f_\mathrm{GC},\alpha_{d}\}$ (\textit{fourth column}) models and the reduced GW dataset. We report posterior medians and 90\% credible intervals.}
   \label{tab:median_err}
\end{table}
The dark blue contours show the marginalised posteriors inferred with the reduced dataset.
All distributions are now unimodal; the corresponding constraints are reported in the third column of Table~\ref{tab:median_err}.

The correlations of~$A_\mathrm{ISO}$ with both~$Z_\mathrm{max}$ and~$f_\mathrm{GC}$ are a consequence of our parametrisation choices.
~$A_\mathrm{ISO}$ correlates with $Z_\mathrm{max}$ because, for instance, a smaller~$A_\mathrm{ISO}$ would imply a decrease of the isolated BBH merger efficiency and, thus, a decrease of the normalisation of the merger rate density.
This must be compensated by a higher metallicity cut-off, in order to keep the expected number of events consistent with the actual observed number.
~$A_\mathrm{ISO}$ and~$f_\mathrm{GC}$ are correlated because~$f_\mathrm{GC}$ is the ratio between the efficiency amplitudes by definition.
The relation between~$Z_\mathrm{max}$ and~$f_\mathrm{GC}$ can be explained in terms of their influence on the mass distribution.
Increasing~$Z_\mathrm{max}$ raises the mass function at the faint end ($<35\ \mathrm{M}_\odot$), which is mainly contributed by isolated BBHs,
while increasing~$f_\mathrm{GC}$ rises it at the bright end ($>35\ \mathrm{M}_\odot$), that is contributed by BBHs from GCs (see~\citealp{Torniamenti2024}).
Both of these increments would then increase the number of expected events.
Moreover, the degeneracies of~$Z_\mathrm{max}$ with~$A_\mathrm{ISO}$ and~$f_\mathrm{GC}$ are alleviated by the fact that both the predicted redshift evolution of merger rate density and mass distribution of isolated BBHs depend on~$Z_\mathrm{max}$.
Thus, fitting for the redshift and mass dependence of the GW events partially reduces these degeneracies.

With the aim of comparing our findings with the reduced dateset to isolated BBH population synthesis results in the literature, we report in Figure~\ref{fig:merger_eff_cp} the median values and relative credible interval obtained from Figure 1 of~\cite{Vanson2025}.
The authors gathered BBH merger efficiencies of multiple studies and calculated the median, 10th and 90th percentiles.
For a qualitative comparison, we take the value of the median at low metallicity as proxy for~$A_\mathrm{ISO}$ and the metallicity at which it decreases by a factor~$10$ as a proxy for~$Z_\mathrm{max}$.
We do the same for the percentiles to get plausible credible intervals.
Firstly, we note that the~$Z_\mathrm{max}$ posterior from the reduced GW dataset, even if in agreement within about~$2\sigma$, peaks at a slightly higher metallicity.
This is due to the metallicity dependence of the isolated BBH mass distribution (Figure~\ref{fig:NBBH_sevn}): a smaller~$Z_\mathrm{max}$ would prevent a proper matching of the~$10\ \mathrm{M}_\odot$ peak.
Secondly, the amplitude of the isolated BBH merger efficiency needs to be a factor~$\sim20$ lower than\ the median found in~\cite{Vanson2025}, in order to fit the number of detected GW events.
This discrepancy is due the fact that, as previously mentioned, current theoretical predictions tend to overestimate the observed merger rate density~\citep{Sgalletta2025,Boco2026a}.
Thirdly, the median value of~$f_\mathrm{GC}$ implies that the GC BBH merger efficiency should be a factor~$\sim30$ larger than that of the isolated channel to reproduce the high-mass range of the mass spectrum.
Hence, the median value of~$A_\mathrm{GC}$ is~$6.46\times10^{-5}\ \mathrm{M}_\odot^{-1}$.
See Section~\ref{sec:sfr_eff} for further discussion.

\begin{figure}
    \centering
    \includegraphics[width=0.9\linewidth]{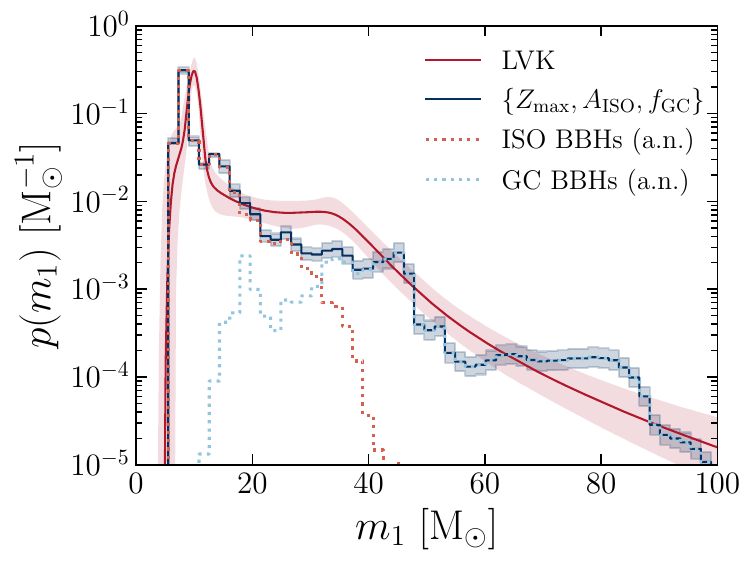}
    \caption{
    BBH primary mass pdf evaluated at~$z=0.2$.
    The dark blue solid curve shows the posterior median for the~$\{Z_\mathrm{max},A_\mathrm{ISO},f_\mathrm{GC}\}$ model and the reduced GW dataset, while the red solid curve the one inferred by the LVK collaboration with the~\textsc{Broken Power Law + 2 Peaks} model~\citep{lvk2026:pop}.
    The dark blue and red shaded regions show the~$90\%$ credible intervals, respectively.
    The dotted red (light blue) line describes the scaled contribution of the isolated (GC) BBH channel.}
    \label{fig:mass_distribution}
\end{figure}

\subsubsection{BBH mass distribution}\label{sec:mass_pdf}
With the posterior samples for the~$\{Z_\mathrm{max},A_\mathrm{ISO},f_\mathrm{GC}\}$ model and the reduced GW dataset, we calculate the BBH primary mass spectrum and merger rate density and compare them with the estimates of the LVK collaboration.
In Figure~\ref{fig:mass_distribution}, we show the inferred primary mass pdf, in dark blue.
This is obtained normalising the total differential merger rate density given by equation~\eqref{eq:mergerRateDensity}, marginalised over the mass ratio.
We report also the result of the LVK collaboration with the~\textsc{Broken Power Law + 2 Peaks} model~\citep{lvk2026:pop} in red.
In addition, we highlight the (scaled) contributions of the different BBH formation channels with dotted curves.
BBH mergers from the isolated channel are able to reproduce the peak around~$10\ \rm M_\odot$~\citep{Torniamenti2024}.
Beyond~$\sim35\ \mathrm{M}_\odot$, the distribution of primary masses is driven by dynamical BBH mergers.
Focusing on the peak around~$\sim35\ \mathrm{M}_\odot$, it is mostly contributed by BBHs in GC but, from our simulations, the peak results to be shifted to slightly higher masses~\citep[see e.g.][]{Farag2022}.
This really depends on the details of both stellar evolution and dynamical pair-up, but it could also be influenced by the metallicity dependence of the star formation activity in GCs.
Moreover, the profile and position of the peak may also vary with contributions of other dynamical channels~\citep[see Figure 5 of][]{Mapelli2022}.
The impact of different pairing functions is explored in Section 4 of~\cite{Torniamenti2024}.

\begin{figure}
    \centering
    \includegraphics[width=0.9\linewidth]{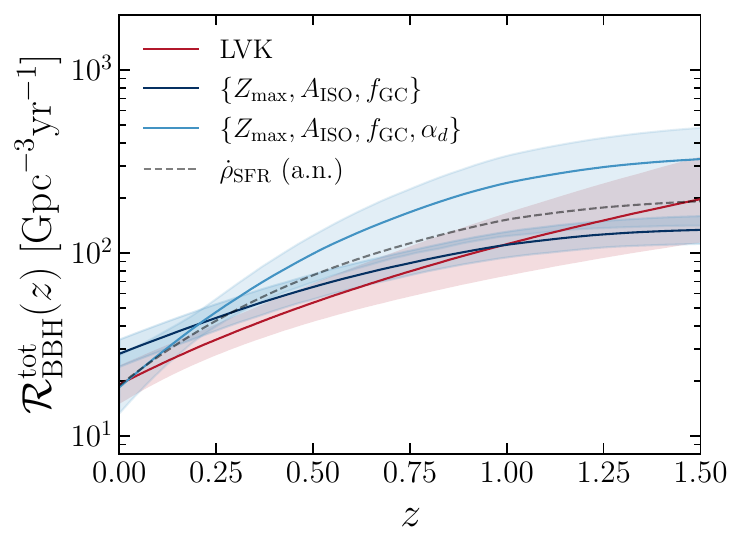}
    \caption{Comparison between the BBH merger rate density inferred with the~$\{Z_\mathrm{max},A_\mathrm{ISO},f_\mathrm{GC}\}$ and~$\{Z_\mathrm{max},A_\mathrm{ISO},f_\mathrm{GC},\alpha_{d}\}$ models, in dark and light blue respectively, and the reduced GW dataset. The estimate from the LVK collaboration with the~\textsc{Power Law Redshift} model~\citep{lvk2026:pop} is in dark red.
    The solid curves show the posterior medians, while the shaded regions the~$90\%$ credible intervals.
    The dashed grey curve describes the scaled cosmic star formation rate density given by equation~\eqref{eq:csfrd}, marginalised over metallicity.}
    \label{fig:merger_rate_density}
\end{figure}

\subsubsection{BBH merger rate density}\label{sec:merger_rate_density}
Figure~\ref{fig:merger_rate_density} shows the BBH merger rate density as a function of redshift.
Although our model successfully reproduces~$N_\mathrm{obs}=256$, one of the key quantities driving the hierarchical Bayesian inference, the inferred local merger rate and slope of the merger rate density do not fully agree with LVK results.
Specifically, our model lies slightly above the LVK inference at~$z\lesssim1$, whereas at~$z\gtrsim1$ its evolution is shallower.
The flattening at~$z\gtrsim1$ is expected as, by construction, we impose a prescribed redshift dependence, anchoring our model on an observation-driven cosmic SFRD, while LVK uses a free parameter for the slope.
From Figure~\ref{fig:merger_rate_density}, it is apparent that in that redshift range the adopted cosmic SFRD evolves more gradually than the~\textsc{Power Law Redshift} model employed, for example, in~\cite{lvk2026:pop}.
The overestimate at~$z\lesssim1$, and consequently of the local merger rate, stems from two main factors.

First, there is a compensation effect arising from the underestimate of the primary mass pdf in correspondence of the~$35\ \mathrm{M}_\odot$ peak: this suppresses the contribution of more massive BBHs, reducing the expected number of detections.
As a consequence, a higher normalization of the merger rate density is required to reproduce~$N_\mathrm{obs}$.
The second aspect is represented by the delay time distribution.
The predicted delay time distributions offset the cosmic SFRD slope, flattening the BBH merger rate density.
This flattening determines an excess of BBH mergers in the local Universe and a shallower slope of the merger rate density profile.
In order to better understand this effect, in the next Section we investigate the impact on our findings of introducing a parametrisation for the delay time pdfs.

Given these considerations, it is not problematic if our inferred merger rate density does not  fully match that inferred by LVK.
Indeed, the two approaches are different: while we assume the redshift and mass dependence by fixing the cosmic SFRD and adopting the mass and delay time distributions from simulations, the LVK population model is fully parametric (at least the one considered in this work), and therefore allows for greater flexibility.

\begin{figure}
    \centering
    \includegraphics[width=0.9\linewidth]{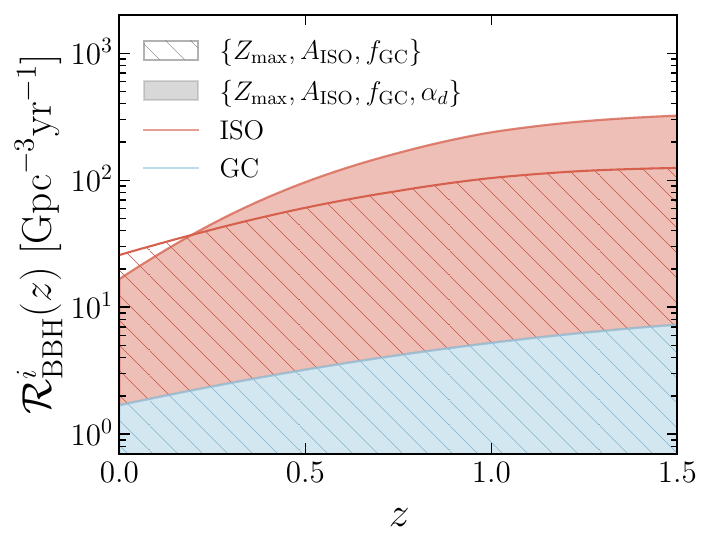}
    \caption{Median BBH merger rate density of the isolated (light red) and GC (light blue) formation channels, inferred with the~$\{Z_\mathrm{max},A_\mathrm{ISO},f_\mathrm{GC}\}$ (hatched shading) and~$\{Z_\mathrm{max},A_\mathrm{ISO},f_\mathrm{GC},\alpha_{d}\}$ model (solid shading), and the reduced GW dataset.}
    \label{fig:merger_rate_density_i}
\end{figure}

Finally, Figure~\ref{fig:merger_rate_density_i} shows the contributions to the total intrinsic merger rate density of the single BBH formation channels computed for this model (hatched shading).
The isolated channel provides the vast majority of the total merger rate density and the redshift evolution of the two contributions recall the cosmic SFRD profile.
The minor contribution of dynamical BBHs is related to the small amount of star formation in GCs, compared to one in the field (see Figure~\ref{fig:csfrd_Z}).
However, when computing the expected number of BBH mergers, which takes into account selection effects of the detectors, the two channels provide comparable contributions:~$\sim129$ of the detected BBH mergers would derive from the isolated formation scenario, while~$\sim127$ are expected to originate from GCs.

\subsection{Impact of varying the delay time distribution}
\label{sec:mrd_model2}
To study the influence of the delay time pdf on the local merger rate and redshift profile of the merger rate density, we replicate the previous analysis parametrising such distribution.
We model the delay time pdf of the isolated channel with a power-law~$p_\mathrm{ISO}(t_d)\propto t_d^{\alpha_d}$, while, for the GC channel, we keep the delay time pdfs from the simulations.
The reason is that in this exploratory analysis we focus on the main drivers of the overall shape and normalisation of the total intrinsic merger rate density, i.e. isolated BBH mergers.
Moreover, it may be noticed that we are neglecting the metallicity dependence of the delay time pdf.
A detailed analysis including a proper metallicity dependence will be part of future developments.

\subsubsection{Interplay between isolated BBH merger efficiency and delay time pdf}\label{sec:eff_ptd}
\begin{figure}
    \centering
    \includegraphics[width=1.0\linewidth]{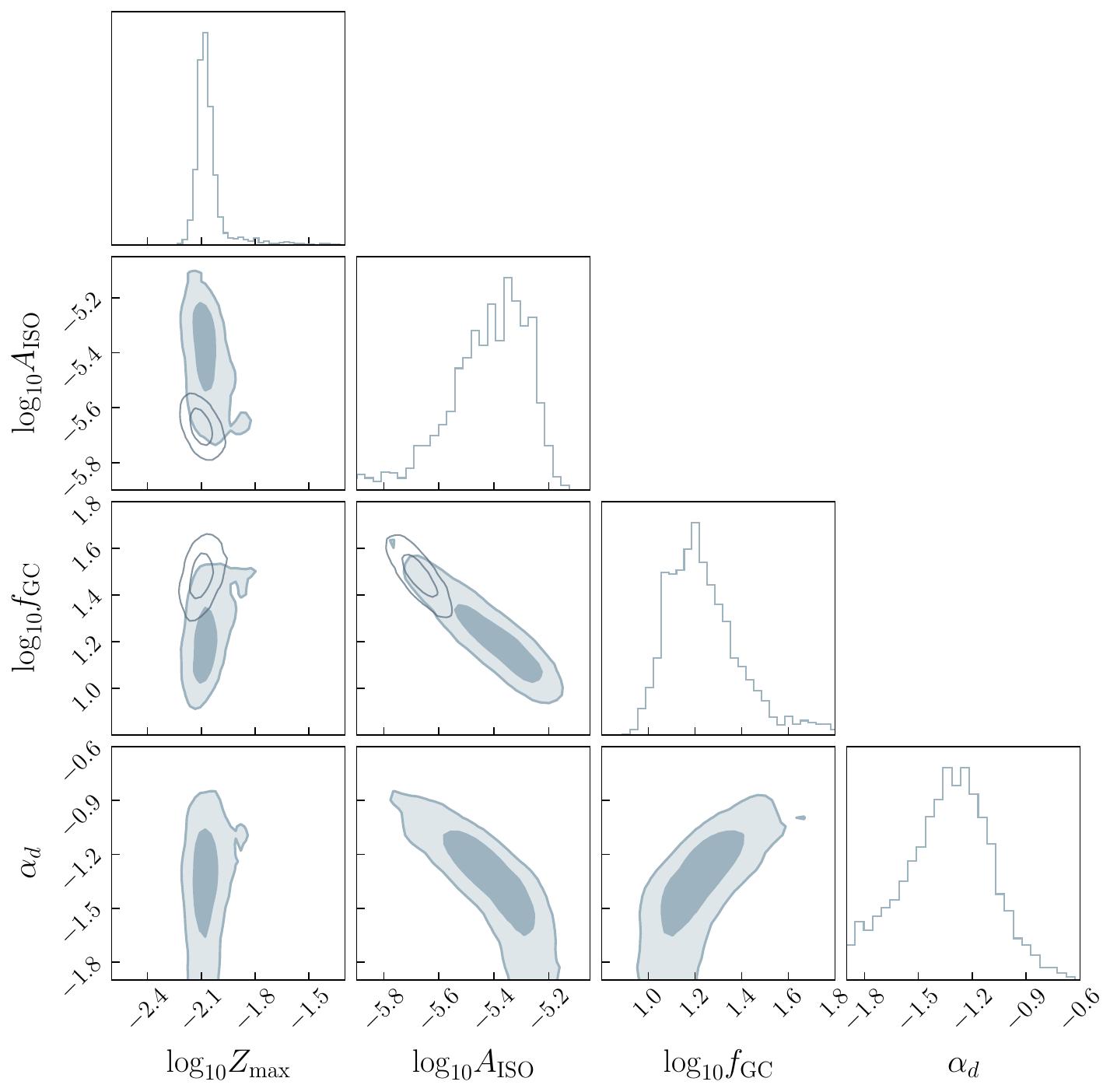}
    \caption{Corner plot with the marginalised posterior distributions obtained from the inference with the~$\{Z_\mathrm{max},A_\mathrm{ISO},f_\mathrm{GC},\alpha_{d}\}$ model and the reduced GW dataset.
    The filled dark (light) blue contour corresponds to the~$50\%$ ($90\%$) credible regions.
    Unfilled contours enclose the~$50\%$ and~$90\%$ credible regions from the~$\{Z_\mathrm{max},A_\mathrm{ISO},f_\mathrm{GC}\}$ model for comparison.}
    \label{fig:merger_eff_td_cp}
\end{figure}

We repeat the hierarchical Bayesian inference on the reduced GW dataset adding~$\alpha_{d}$ as a new free parameter.
We report in the fourth column of table~\ref{tab:median_err} the constraints and, in Figure~\ref{fig:merger_eff_td_cp}, the new set of marginalised distributions obtained with the~$\{Z_\mathrm{max},A_\mathrm{ISO},f_\mathrm{GC},\alpha_{d}\}$ model.

The anti-correlation between~$\alpha_{d}$ and~$A_\mathrm{ISO}$ is evident.
Smaller exponents of the delay time pdf favor shorter delay times and, thereby, a reduced fraction of BBHs in the local Universe~\citep[see e.g. Figure 11 of][]{Santoliquido2021}.
Thus, larger values of $A_\mathrm{ISO}$ are required to maintain consistency between the expected and observed number of mergers.
The correlations of~$\alpha_{d}$ with both~$Z_\mathrm{max}$ and~$f_\mathrm{GC}$ can be traced back to their interplay with~$A_\mathrm{ISO}$, while the correlations among~$Z_\mathrm{max}$,~$A_\mathrm{ISO}$ and~$f_\mathrm{GC}$ remain similar to the previous model.

It may be noticed that, although the inclusion of~$\alpha_{d}$ leaves the posterior of~$Z_\mathrm{max}$ nearly unchanged, the peak of the posterior of~$f_\mathrm{GC}$ shifts toward lower values.
This behavior is driven by the larger~$A_\mathrm{ISO}$ favored by smaller delay time exponents.
The median value of~$A_\mathrm{ISO}$ is now roughly~$10$ times lower the median of~\cite{Vanson2025}, so the difference has been halved compared to the previous estimate.
Consequently, the updated median value of~$f_\mathrm{GC}$ indicates that the GC BBH merger efficiency needs to be~$\sim16$ times larger than~$A_\mathrm{ISO}$.
This result is broadly in agreement with the simulations of~\cite{Sedda2026:bpop}~\footnote{The data are publicly available at~\cite{Sedda2026:bpop_zenodo}.}, where the ratio between the merger efficiencies of BBH in GCs and isolation at low metallicity is~$\sim10$, matching the lower tail of our~$\mathrm{log}_{10}f_\mathrm{GC}$ posterior.
Finally, because of the lower~$f_\mathrm{GC}$, the shape of the primary mass pdf mildly varies, becoming lower in the~$[20,35]\ \mathrm{M}_\odot$ range, as illustrated in Figure~\ref{fig:mass_distribution_comparison} in Appendix.

\subsubsection{BBH merger rate density}\label{sec:merger_rate_density_1}
Isolated BBH evolution usually predicts power-law delay time distribution with an exponent~$\alpha_{d}\simeq-1$, as a consequence of the distribution of the BBH semi-major axis at formation~\citep{Sana2012}.
Anyway, this may vary appreciably depending on metallicity~\citep{Lamberts2016} and processes such as natal kicks and mass transfer~\citep{Mapelli2017}.
The light blue curve in Figure~\ref{fig:merger_rate_density}, derived from our analysis and corresponding to~$\alpha_{d}\simeq-1.33$, suggests that a steeper distribution would better reconcile the local merger rate with the LVK estimate~\citep{Boco2026a}.
Given the steeper delay time distribution,~$A_\mathrm{ISO}$ moves to larger values, as shown in table~\ref{tab:median_err}, so that the~$N_\mathrm{obs}$ constraint is still respected.

The solid shaded areas in Figure~\ref{fig:merger_rate_density_i} illustrate how the intrinsic merger rate densities from the single channels might vary accordingly to the new parametrisation: the GC contribution is left unchanged, whereas the slope of isolated contribution increases thanks to the inferred delay time pdf.
As predictable, the relative contributions in terms of the expected numbers of events is basically left unchanged.

\begin{figure}
    \centering\includegraphics[width=0.9\linewidth]{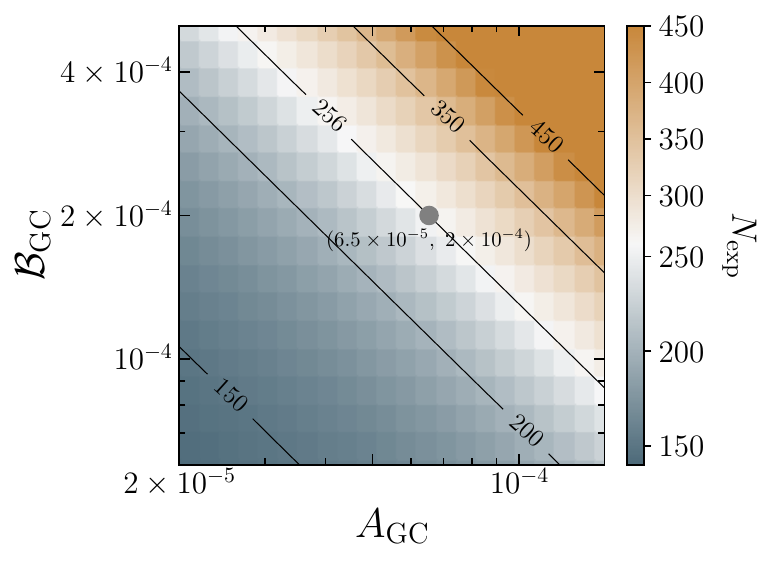}
     \caption{Variation of the expected number of BBH mergers in terms of the amplitude of the merger efficiency of BBHs in GCs,~$A_\mathrm{GC}$, and the normalisation of the cosmic star formation rate density in GCs,~$\mathcal{B}_\mathrm{GC}$, for the~$\{Z_\mathrm{max},A_\mathrm{ISO},f_\mathrm{GC}\}$ model and the reduced GW dataset.
     The black contours show levels of constant expected number ranging from~$150$ to~$450$.
     The grey disk denotes~$N_\mathrm{exp}$ computed for the parameter maximum a posteriori values and fiducial~$\mathcal{B}_\mathrm{GC}$.
     Its coordinates in the plane are reported in parenthesis.}
     \label{fig:Nexp_Bgc_Agc}
 \end{figure}

\section{Discussion}\label{sec:discussion}

\subsection{Degeneracy between star formation in GCs and BBH merger efficiency}\label{sec:sfr_eff}
The star formation rate density (SFRD) in star clusters is affected by strong uncertainties.
In our case, the star formation activity that occurs in globular clusters is described by equation~\ref{eq:csfrd_gc}.
The fiducial normalisation factor is~$\mathcal{B}_\mathrm{GC}=2\times10^{-4}\ \mathrm{M}_\odot\mathrm{Mpc}^{-3}\mathrm{yr}^{-1}$~\citep{Mapelli2022,Torniamenti2024}.
Such value is assumed to be consistent with observational constraints of globular clusters in the Milky Way \citep{ReinaCampos2019:gc} and previous numerical models \citep{rodriguez2018,elbadry2019}.
Variations in the normalization factor would reflect in changes of the merger rate density, as discussed in Section 4 of~\cite{Mapelli2022}.
Consequently, they would also alter the inference of the merger efficiency parameters.
Equation~\ref{eq:mergerRateDensity_i} indicates that the merger rate density of BBHs in GCs is directly proportional to the product of~$A_\mathrm{GC}$ and~$\mathcal{B}_\mathrm{GC}$, which is the actual quantity we can constrain.

In Figure~\ref{fig:Nexp_Bgc_Agc}, the degeneracy between~$A_\mathrm{GC}$ and~$\mathcal{B}_\mathrm{GC}$ is illustrated in terms of the expected number of BBH mergers for the~$\{Z_\mathrm{max},A_\mathrm{ISO},f_\mathrm{GC}\}$ model.
~$\mathrm{log}_{10}Z_\mathrm{max}$ and~$\mathrm{log}_{10}A_\mathrm{ISO}$ are fixed at their maximum a posteriori values, which are~$-2.10$ and~$-5.67$, respectively.
The black isocontours corresponding to~$N_\mathrm{exp}=256$ show how changes in the normalisation factor of the cosmic SFRD in GCs are compensated by different efficiency amplitudes, so that~$N_\mathrm{exp}$ can be in agreement with~$N_\mathrm{obs}=256$.

Hence, the constraint on~$A_\mathrm{GC}$ is a consequence of the assumed~$\mathcal{B}_\mathrm{GC}$, which is largely uncertain.
The median value of~$A_\mathrm{GC}$ for the~$\{Z_\mathrm{max},A_\mathrm{ISO},f_\mathrm{GC}\}$ model is~$\sim6.46\times10^{-5}\ \mathrm{M}_\odot^{-1}$, for our fiducial~$\mathcal{B}_\mathrm{GC}$.
This is a factor~$\sim7$ lower than the BBH merger efficiency in GCs at low metallicity estimated in~\cite{Sedda2026:bpop}.
If we assume their fiducial value for~$\mathcal{B}_\mathrm{GC}$,~$1.2\times10^{-4}\ \mathrm{M}_\odot\mathrm{Mpc}^{-3}\mathrm{yr}^{-1}$, we would infer an higher~$A_\mathrm{GC}$, reducing the difference by about a factor 2.
This applies also for the~$\{Z_\mathrm{max},A_\mathrm{ISO},f_\mathrm{GC},\alpha_{d}\}$ model, since the inferred median of~$A_\mathrm{GC}$ minutely varies.

\subsection{Intrinsic BBH merger rate vs. observed events}\label{sec:R0_Nobs}

Since our goal is to find the values of the BBH merger efficiency, in isolation and in globular clusters to reproduce the observed merger rates, one might wonder why the inferred models does not  always match the LVK merger rate density in the local Universe or at higher redshift (see Figure~\ref{fig:merger_rate_density}).
This is due to the fact that our hierarchical Bayesian framework does not aim to reproduce intrinsic properties of the BBH population, like the intrinsic merger rate or the mass distribution, derived by LVK.
Rather, it infers the model parameters by simultaneously fitting the observed number of events and the detected masses and redshift of single events.
The intrinsic properties of the population are then reconstructed a posteriori.
This is the same procedure adopted by the LVK collaboration, where the mass function and merger rate density are modeled with a parametric or agnostic approach and then fitted to the observations.
Therefore, the two inferences should be considered as two distinct modeling approaches applied to the same underlying data, i.e. the number of detected GW events, their observed masses and redshifts.
In other words, we are not trying to reproduce the intrinsic population properties obtained by LVK; instead, we are directly fitting the data with another model, where the BBH merger efficiencies in isolation and in globular clusters act as free parameters.
Small differences in the reconstructed population properties are thus natural and expected, as discussed in Sections~\ref{sec:merger_rate_density} and~\ref{sec:merger_rate_density_1}.

\section{Conclusions}\label{sec:conclusions}
In this work, we built a semi-parametric BBH population model for both isolated and dynamically-formed BBHs.
Starting from an observational metallicity-dependent cosmic SFRD, we parametrise the merger efficiency of BBHs in isolation and in globular clusters, deriving their values by fitting the model to GW events from the GWTC-5.0, within a full hierarchical Bayesian framework.
Specifically, we constrain the merger efficiencies necessary to simultaneously reproduce the number of detected GW events, alongside their redshift and mass distributions.
By applying this inference framework, we find that:
\begin{itemize}[label=\textbullet]
    \item The constrained isolated BBH efficiency features a sharp drop at~$\mathrm{log}_{10}Z_\mathrm{max}\sim-2.10$, which is consistent within about~$2\sigma$ with populations synthesis results gathered in~\cite{Vanson2025}.
    \item The amplitude of the isolated BBH merger efficiency should be suppressed by a factor~$\sim20$ with respect standard findings of population synthesis codes.
    Such a reduction would significantly helps in reconciling the predicted merger rate density with GW observations.
    \item The globular cluster formation scenario requires an efficiency~$\sim30$ times larger than that of isolated BBHs to properly fit the high-mass tail of the BBH mass distribution.
    \item While the isolated formation channel dominates the total intrinsic merger rate density, both formation channels contribute comparably to the expected number of events. Indeed,~$\sim129$ of the detected BBH mergers are predicted to originate from the isolated formation scenario, whereas~$\sim127$ are expected to derive from the dynamical channel.
\end{itemize}

In Section~\ref{sec:mrd_model2}, we introduce a power-law parametrisation of the isolated BBH delay time pdf.
Employing the same inference procedure yields the following:
\begin{itemize}[label=\textbullet]
    \item There is a clear degeneracy between the power-law exponent,~$\alpha_{d}$, and the amplitude of the isolated channel efficiency,~$A_\mathrm{ISO}$.
    Smaller values of~$\alpha_{d}$ favor a reduced fraction of BBH in the local Universe; therefore, a larger~$A_\mathrm{ISO}$ is required to maintain consistency between the predicted and observed number of GW events.
    \item The peak of the~$A_\mathrm{ISO}$ posterior shifts to higher values, halving the reduction effect found with the previous parametrisation.
    Thus, the inferred median value of~$f_\mathrm{GC}$ is reduced by a factor 2, meaning that the GC BBH merger efficiency needs to be~$\sim16$ times larger than~$A_\mathrm{ISO}$ in this case.
    \item Given our assumed cosmic SFRD, a delay time distribution~$p_\mathrm{ISO}(t_d)\propto t_d^{\alpha_d}$ with~$\alpha_{d}\simeq-1.33$ provides a steeper merger rate density that matches the LVK estimate at~$z\simeq0$.
\end{itemize}

We finally show that the inferred amplitude of the merger efficiency of BBHs in globular clusters,~$A_\mathrm{GC}$, is highly degenerate with the assumed normalisation factor of the cosmic SFRD in GCs,~$\mathcal{B}_\mathrm{GC}$.
Consequently, if for instance a larger~$\mathcal{B}_\mathrm{GC}$ is considered, the constraint on~$A_\mathrm{GC}$ scales down by the same factor, ensuring that the population model consistently fits GW observations.

Overall, our analysis identifies the merger efficiency and delay time distribution as the primary factors to address in order to alleviate the main tension between theoretical predictions of the BBH merger rate density and GW observations, providing quantitative benchmarks to guide future binary evolution studies.
In future work, we will further investigate how the merger efficiency is affected by the delay time distribution, focusing on the main binary evolution processes responsible for shaping the delay time distribution itself.
We will also include additional dynamical channels adopting other population synthesis codes to obtain unbiased constraints through a more accurate modeling of the BBH mass spectrum.
Lastly, the inclusion of the third part of the fourth observing run, together with upcoming observations, will provide increasingly stringent tests of our framework, helping to bridge the gap between BBH population synthesis predictions and the observed GW population.

\begin{acknowledgements}
MB thanks Giovanni Antinozzi for helpful discussions. MB acknowledges that this article was produced while attending the PhD program in PhD in Space Science and Technology at the University of Trento, Cycle XXXIX, with the support of a scholarship financed by the Ministerial Decree no. 118 of 2nd March 2023, based on the NRRP - funded by the European Union - NextGenerationEU - Mission 4 "Education and Research", Component 1 "Enhancement of the offer of educational services: from nurseries to universities” - Investment 4.1 “Extension of the number of research doctorates and innovative doctorates for public administration and cultural heritage” - CUP E66E23000110001 and support by the Italian grant Project SPACE-IT-UP by the Italian Space Agency and Ministry of University and Research, Contract Number 2024-5-E.0. LB acknowledges support by the Deutsche Forschungsgemeinschaft (DFG, German Research Foundation) in the form of a Walter Benjamin position -- Projektnummer 555003977. SR acknowledges funding by the Deut\-sche For\-schungs\-ge\-mein\-schaft – project number 546677095. CS acknowledges financial support from the Alexander von Humboldt Foundation for the Humboldt Research Fellowship. AL has been supported by the Istituto Nazionale di Fisica Nucleare (INFN) via the specific national initiative QGSKY.
The authors acknowledge financial support from the German Excellence Strategy via the Heidelberg Cluster of Excellence (EXC 2181 - 390900948) STRUCTURES.
The authors acknowledge support by the state of Baden-W\"urttemberg through bwHPC, the German Research Foundation (DFG) through grants INST 35/1597-1 FUGG and INST 35/1503-1 FUGG and also from the European Research Council for the ERC Consolidator grant DEMOBLACK, under contract no. 770017 (PI: M. Mapelli) and for the ERC Advanced grant IMBLACK, under contract no. 101197608 (PI: M. Mapelli).
\end{acknowledgements}

\bibliographystyle{aa}
\bibliography{references}

\appendix

\section{Gravitational wave data}\label{sec:appendix_gw_data}
In our analysis we consider BBH events from the GWTC-5.0 catalog~\citep{Abac2026:o4_2}.
More in detail, we start with selecting 259 BBH mergers with the false alarm rate (FAR)~$<1\ \mathrm{yr}^{-1}$ as threshold for detection, following~\cite{lvk2026:pop}, and then we consider a subsample of 256 events as explained in~\ref{sec:mrd_model1}.

For computing the integral in equation~\eqref{eq:likelihood}, we use the posterior samples from the parameter estimation (PE) of individual binaries reported in~\citep{lvk2022:gwtc2.1_pe,lvk2023:gwtc3_pe,lvk2026:gwtc41_pe,lvk2026:gwtc51_pe,lvk2026:gwtc52_pe}, depending on the observing run during which the specific events were detected.
Moreover, to account for selection biases in the calculation of~$N_\mathrm{exp}$, we exploit the injections used for the cumulative LVK search sensitivity estimates~\citep{Essick2025, lvk2026:gwtc5_inj}.
They contain both semi-analytic injections for the first two runs, O1 and O2~\citep{Essick2023}, and real-noise injections for O3 and the first two parts of the fourth observing run, O4a and O4b.

\section{Likelihood evaluation}\label{sec:appendix_likelihood}
GW event posterior samples and injections are needed to evaluate the hierarchical likelihood via Monte Carlo integration, as performed by~\textsc{icarogw}~\citep{Mastrogiovanni2023:icarogw}.
In particular, we use~$\sim3\times10^3$ posterior samples to evaluate the integrand of equation~\ref{eq:likelihood} for each event, and~$\mathcal{O}(10^6)$ injections to compute~$N_\mathrm{exp}$ with the detector selection effects.

In order then to calculate the likelihood in the parameter space of our population model, we use the nested sampler~\textsc{dynesty}~\citep{Speagle2020:dynesty}, which is in the~\textsc{bilby} Bayesian inference package~\citep{Ashton2019:bilby,Smith2020:bilby,Romero-Shaw2020:bilby}.
Finally, for numerical stability reasons, we set a
maximum variance of~$1$ for the hierarchical likelihood~\citep{Talbot2023,lvk2026:pop}.

\section{BBH catalogs}\label{sec:appendix_bbh_cat}

\subsection{Isolated BBHs}\label{sec:sevn}

\begin{figure*}
     \centering\includegraphics[width=0.8\linewidth]{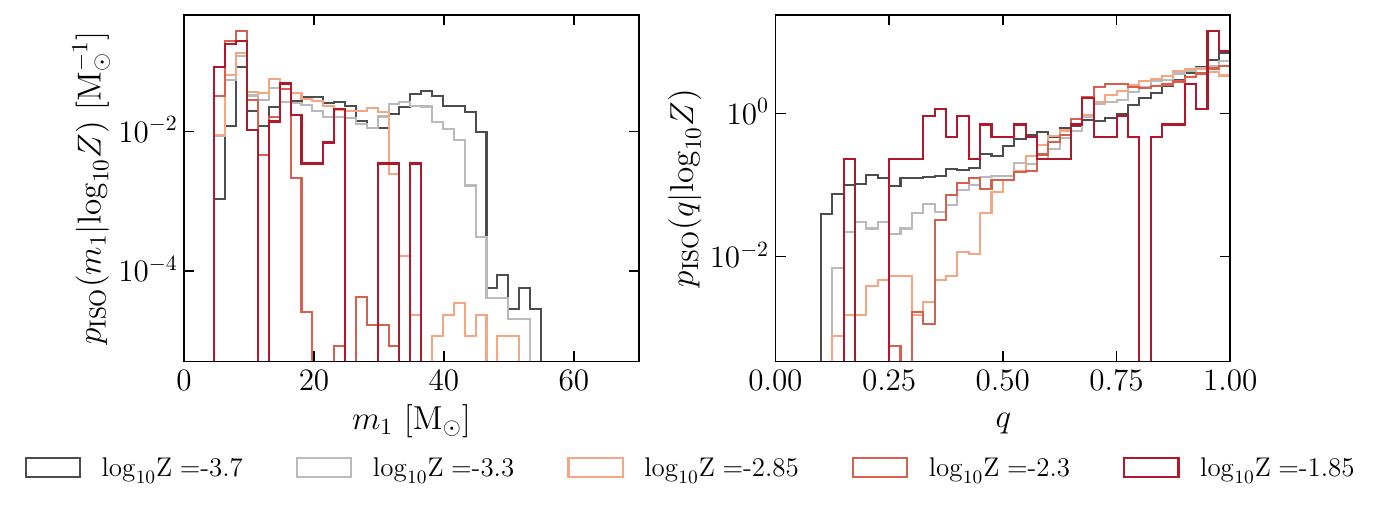}
     \caption{Primary mass (\textit{left panel}) and mass ratio (\textit{right panel}) pdfs of isolated BBHs that merge within a Hubble time, at given metallicity values.}
     \label{fig:NBBH_sevn}
 \end{figure*}

We use~\textsc{sevn}~\citep{Spera2017, Spera2019, Mapelli2020, Iorio2023} to evolve a population of isolated stellar binaries in the field.
We generate the initial conditions according to~\cite{Sgalletta2025} and we adopt the fiducial setup of~\cite{Iorio2023}.
Specifically, we assume the~\textit{rapid} model for core-collapse supernovae by~\cite{Fryer2012}, the pair-instability treatment of~\cite{Mapelli2020b} and the common envelope efficiency~$\alpha_\mathrm{CE}=1$.
Following~\cite{Giacobbo2020}, we draw the natal kicks from a Maxwellian distribution with~$\sigma=265\ \mathrm{kms}^{-1}$~\citep{Hobbs2005}.
We rescale them by a factor~$\propto M_{\rm ej}/ M_{\rm rem}$, where~$M_{\rm ej}$ and~$M_{\rm rem}$ are, respectively, the mass of the ejecta and compact remnant.
Note that we expect the corrections introduced by~\cite{Disberg2025} to have a minimal impact on the BBH merger rate since fallback is important for the vast majority of the population and because of our parametric approach.
We simulate~$10^7$ binaries for~$15$ metallicities:
$Z_\mathrm{sim}=0.0002,0.0003,0.0004,0.0005,0.0007,0.001,0.0014,0.002,\\
0.003,0.004,0.005,0.007,0.01,0.014,\text{and}\ 0.02$.
We obtain therfore~$15$ catalogs of isolated BBHs.
The total simulated stellar mass for each metallicity is $M^\mathrm{sim}_\star = 2.2\times10^8\ \mathrm{M}_\odot$.

In Figure~\ref{fig:NBBH_sevn} we show the resulting normalised distributions for the primary mass and mass ratio of those isolated BBHs that merge within a Hubble time, at different metallicity values.
In addition to impacting the number of systems that actually merge, metallicity also influences both the maximum BH mass and the peak around~$10\ \mathrm{M}_\odot$, due to the effect of stellar winds~\citep{Iorio2023}.
This metallicity dependency is then inherited by the mass ratio distribution.

\subsection{BBHs in globular clusters}\label{sec:fastcluster}
We model the formation of dynamical and hierarchical BBH mergers in globular clusters with~\textsc{fastcluster}~\citep{Mapelli2021, Mapelli2022, Vaccaro2023, Torniamenti2024}.
We sample the initial BH populations from the same~\textsc{sevn} catalogs as for the isolated formation channel (see Appendix~\ref{sec:sevn}).
We model the BBH dynamical pairing following the sampling criterion introduced in~\cite{Antonini2023}, which preferentially couples the most massive objects within the cluster core~\citep{Heggie1975}.
We draw the initial cluster mass from a Gaussian distribution with mean~$\langle\mathrm{log}_{10} M_\mathrm{tot}/M_\odot\rangle=5.9$, assuming a fiducial standard deviation~$\sigma_{\mathrm{log}_{10}M_\mathrm{tot}}=0.4$.
The GC density at half-mass radius is sampled again from a Gaussian distribution with mean~$\langle\mathrm{log}_{10} \rho/(M_\odot\mathrm{pc}^{-3})\rangle=4.7$, and fiducial standard deviation~$\sigma_{\mathrm{log}_{10}\rho}=0.4$ (see Appendix B of~\citealt{Torniamenti2024}).
Moreover, we adopt the same core-collapse supernovae, pair-instability, natal kick models and common envelope efficiency employed in the~\textsc{sevn} runs.
Finally, the BH spin magnitudes are drawn from a Maxwellian distribution with~$\sigma_\chi=0.1$, as done in~\cite{Torniamenti2024}.
We simulate~$10^6$ BBHs for the same~$15$ input metallicities of~\textsc{sevn}.

\section{Cosmic star formation}\label{sec:appendix_csfrd}
Here we detail the distributions involved in the cosmic SFRD calculation (see Equation~\ref{eq:csfrd}):
\begin{itemize}[label=\textbullet]
    \item The star-forming galaxy stellar mass function is the one obtained in~\cite{Chruslinska2019}; it provides an average fit to different observational studies~\citep[e.g.][]{Ilbert2013,Muzzin2013, Tomczak2014, Davidzon2017}.

    \item The SFR pdf is a double Gaussian in~$\mathrm{log_{10}SFR}$, with the two components describing two separate galaxy populations~\citep{Sargent2012, Schreiber2015, Boco2021}.
    The first and most abundant component is constituted by main-sequence (MS) galaxies, i.e. galaxies satisfying a well-known correlation in the~$\mathrm{log_{10}SFR}-\mathrm{log}_{10}M_\star$ plane~\citep[e.g.][]{Speagle2014, Rodighiero2015, Mancuso2016b, Popesso2023}.
    The second peak represents starburst (SB) galaxies, i.e. galaxies that experience a more intense star formation and lie above the main sequence by 0.5-1 dex~\citep[e.g.][]{Rodighiero2011, Bisigello2018, Rinaldi2022, Rinaldi2025}.
    We employ the MS of~\cite{Popesso2023}, since it is the most recent determination, spanning a huge range in mass and redshift, with a scatter~$\sigma_\mathrm{MS}=0.188$ dex~\citep{Sargent2012,Boco2021}.
    The starburst sequence is set to be~$\sim0.6$ dex above the main sequence with a scatter~$\sigma_\mathrm{SB}=0.243$ dex~\citep{Sargent2012,Boco2021}.
    As for the relative abundance of starbursts with respect to main sequence galaxies, we adopt the fraction of starbursts,~$f_\mathrm{SB}$, determined in~\citep{Chruslinska2025}, who use the results of recent observational determinations~\citep{Caputi2017,Bisigello2018,Rinaldi2022,Rinaldi2025}.

    \item Following~\cite{Boco2021}, we model the metallicity pdf using a Gaussian distribution in~$\mathrm{log}_{10}Z$, with the mean determined from the fundamental metallicity relation (FMR).
    This relation links metallicity to both stellar mass and SFR~\citep[e.g.][]{Mannucci2010,Mannucci2011,Andrews2013,Curti2020,Curti2023,Nakajima2023}.
    The FMR is parametrised as in~\cite{Chruslinska2021} and \citet{Boco2026a}. This is a flexible parametrization which inherits the functional form of \citet{Curti2020}, with some free parameters that allow to vary its shape and normalization, thus encompassing several metallicity determinations of different studies. At low stellar mass the relation reduces to $\log Z\propto\gamma\,(\log M_\star-\alpha\,\log \textrm{SFR})$. We choose $\gamma=0.43$ and $\alpha=0.66$ to be in agreement with the FMR found by \citet{Andrews2013}, which has been proven to hold up to $z\sim 6-7$ \citep{Nakajima2023}. At high stellar mass, the relation saturates to a value, $\log Z_0$. We select $\log Z_0=9$, following \citet{Chruslinska2025}, which is a good compromise between estimates from direct method \citep{Pettini2004, Curti2020, Sanders2021} and theoretical models \citep{Kewley2002, Tremonti2004, Kobulnicky2004, Mannucci2010}. For more details on the parameter choice see \citep{Chruslinska2021, Boco2026a}.

    Note that the FMR represents the oxygen abundance in galaxies, since gas metallicity measurements mostly rely on oxygen lines.
    However, binary evolution is mostly affected by the abundance of iron-group elements, since line-driven winds are powered by the iron bump.
    Studies computing BBH merger rates usually rescale oxygen abundance to total metallicity assuming solar relative abundances.
    This assumption might create biases in the estimation of the BBH merger rate density, as recently shown by~\cite{Chruslinska2025} and~\cite{Boco2026a}.
    This is because iron and oxygen grow on different timescale inside galaxies, with many high-$z$ star-forming galaxies being alpha-enhanced~\citep{Boco2026a}.
    To remove this bias we correct the FMR by using the relation between~$[\mathrm{O/Fe}]\equiv\mathrm{log}_{10}(\mathrm{O/Fe})-\mathrm{log}_{10}(\mathrm{O/Fe})_\odot$ and specific SFR ($\textrm{sSFR}\equiv\mathrm{SFR}/M_\star$) derived by~\citep{Chruslinska2024,Chruslinska2025}.
    More details can be found in~\cite{Boco2026a}, where such rescaling procedure has been exploited for the BBH merger rate calculation for the very first time.
\end{itemize}

\section{BBH mass distribution with the~$\{Z_\mathrm{max},A_\mathrm{ISO},f_\mathrm{GC},\alpha_{d}\}$ model}\label{sec:appendix_model_td}

Here we present supplementary results for the model where the delay time pdf of the isolated BBH channel is parametrised with a power-law~$p_\mathrm{iso}(t_d)\propto t_d^{\alpha_d}$.

\begin{figure}
    \centering
    \includegraphics[width=0.9\linewidth]{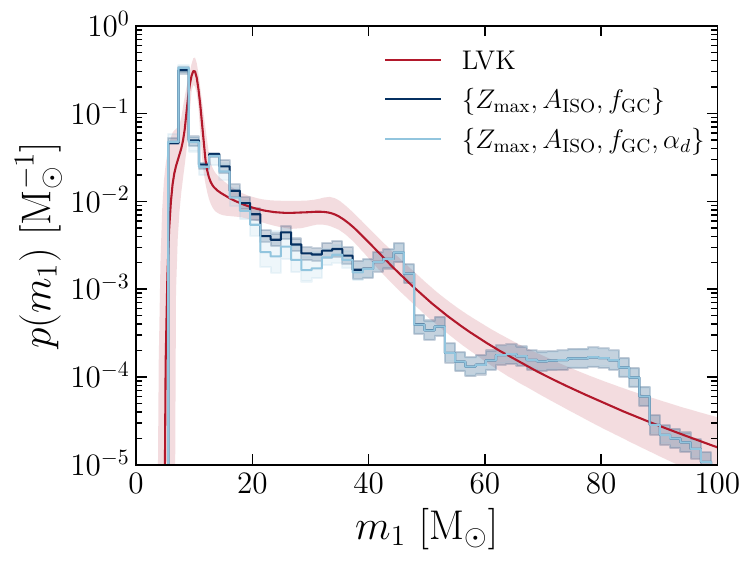}
    \caption{Comparison among BBH primary mass pdfs (evaluated at~$z=0.2$).
    The dark (light) blue solid curve shows the posterior median for the~$\{Z_\mathrm{max},A_\mathrm{ISO},f_\mathrm{GC}\}$ ($\{Z_\mathrm{max},A_\mathrm{ISO},f_\mathrm{GC},\alpha_{d}\}$) model obtained with the reduced GW dataset. The red solid curve corresponds the pdf inferred by the LVK collaboration with the~\textsc{Broken Power Law + 2 Peaks} model~\citep{lvk2026:pop}.
    The shaded regions show the~$90\%$ credible intervals.}
    \label{fig:mass_distribution_comparison}
\end{figure}

In Figure~\ref{fig:mass_distribution_comparison} the BBH primary mass distribution inferred with this model is compared to that derived with the~$\{Z_\mathrm{max},A_\mathrm{ISO},f_\mathrm{GC}\}$ parametrisation and by LVK.
As discussed in Section~\ref{sec:mrd_model2}, the inferred median delay time exponent,~$\alpha_{d}\simeq-1.33$, yields~$A_\mathrm{ISO}$ to a larger value with respect to the~$\{Z_\mathrm{max},A_\mathrm{ISO},f_\mathrm{GC}\}$ model.
This is because of the anti-correlation between~$\alpha_{d}$ and~$A_\mathrm{ISO}$.
As a consequence,~$f_\mathrm{GC}$ decreases and the shape of the primary mass pdf becomes slightly lower in the~$[20,35]\ \mathrm{M}_\odot$ range.

\end{document}